\documentclass[12pt]{iopart}

\begin{document}

\title{Cosmological Magnetic Fields}

\author{Kerstin E. Kunze}

\address{ Departamento de F\'\i sica Fundamental and {\sl IUFFyM},
Universidad de Salamanca,
Plaza de la Merced s/n, 37008 Salamanca, Spain}
\ead{kkunze@usal.es}
\begin{abstract}
Magnetic fields are observed on nearly all scales in the universe, from stars and galaxies upto galaxy clusters and even beyond. 
The origin of cosmic magnetic fields is still an open question, however a large class of models puts its origin in the very early universe.
A magnetic dynamo amplifying an initial seed magnetic field could explain the present day strength of the galactic magnetic field. 
However, it is still an open problem how and when this initial magnetic field was created. 

Observations of the cosmic microwave background (CMB) provide a window to the early universe and might therefore be able to tell 
us whether cosmic magnetic fields are of primordial, cosmological origin and at the same time constrain its parameters. 

We will give an overview of the observational evidence of large scale magnetic fields,  describe generation mechanisms of 
primordial magnetic fields and possible imprints in the CMB.
\end{abstract}

\maketitle

\section{Introduction}

There is evidence for magnetic fields on small upto very large scales. Going beyond stars there are observations of magnetic fields
in the interstellar medium of  galaxies and cluster of galaxies. 

Important tracers of galactic and extragalactic magnetic fields are diffuse synchrotron radiation and  Faraday rotation.
Synchrotron radiation is emitted by electrons spiralling around the magnetic field lines and the emissivity is
determined by the energy spectrum and number density of the electrons, the frequency and the component 
of the magnetic field perpendicular to the line of sight. 
Due to its high degree of intrinsic linear polarization the polarization of the diffuse synchrotron radiation
can be used to determine the structure of the magnetic field. Its degree of polarization is determined by the 
spectral index of the energy spectrum of the electrons. For example, in the Milky Way the degree of polarization in 
a homogeneous magnetic field is 75\% \cite{cmf}. A lower degree of polarization could indicate inhomogeneities in the
magnetic field or the electron distribution.
The magnetic field component along the line of sight, $B_{\parallel}$,  determines the Faraday effect. The polarization plane of a linearly polarized wave 
with wave length $\lambda$ passing through a magnetized medium  is rotated by an angle $RM \lambda^2$ where the rotation measure $RM$ is given by the 
intgral over the path length,
$RM\propto\int n_e B_{\parallel} ds$ (rad m$^{-2}$) where  $n_e$ is the electron density \cite{zh,cmf}.
Unless there is independent information about the electron energy distribution the magnetic strength is determined
assuming equipartition. This means equalizing the energy densities in the magnetic field and electrons.
In our own galaxy this hypothesis can be tested. It is found to be in good agreement with magnetic field estimates using independent
information of cosmic ray energy distributions \cite{beck}. 
Zeeman splitting of spectral lines is another possibility though generally it is very limited due to the much larger line width \cite{zh}. 

Over recent years methods are being developed to detect truly cosmologically magnetic fields, not associated with any virialized structure.
The spectral energy distribution of  some TeV  blazars in the TeV and GeV range hint at the presence of a cosmological
magnetic field pervading all space \cite{vovk}. TeV blazars are a type of active galactic nucleus (AGN) which produce $\gamma$ ray photons in the
TeV energy range. These photons cannot travel very far from the source since they interact with the extragalactic background light producing electron positron pairs. These particles interact with the photons of the cosmic microwave background (CMB)  via inverse Compton effect thereby emitting secondary photons in the 
GeV energy range. Thus an electromagnetic cascade takes place. The trajectories of the  electrons and positrons in this cascade 
are deflected if the cascade takes place in a magnetized medium. This could then lead to time-delayed observation of the GeV signal, 
the detection of an extended emission of an initially point-like source or the absence of power in the GeV part of the energy spectrum of the
source since the charged particles have been deflected out of our line of sight.
Observations of any of these effects imply a lower bound on any cosmologically magnetic field. An interesting example
is the TeV Blazar 1ES0229+200. From the observations of HESS and VERITAS telescopes indicate no 
significant time variability in the TeV energy flux over the time scale of observation of 3 yr. 
The corresponding  Fermi/LAT data of this source show
a lack of power in the GeV energy range. Depending on the  particular model of the geometry and 
parameters of the electromagnetic cascades the lower limit of the intergalactic 
magnetic field is estimated to be larger than $B>5\times 10^{-15}$ G \cite{tav} or $B>10^{-18}$ G \cite{der}.
AGNs are thought to produce cosmic rays, in particular protons. These interact with the CMB photons as well as the EBL photons and produce
secondary high energy photons at energies above TeV \cite{ekkb}.
Including the cosmic ray contribution solves certain
puzzles with the TeV data such as lack of time correlations and leads to a limit of $1\times 10^{-17}<B<10^{-14}$ G \cite{eak}.

Magnetic fields play an important role in  the physics 
of star formation as they allow to reduce angular momentum of the protostellar cloud during collapse.
Moreover, the magnetic pressure acts against gravitational collapse. Thus the presence of 
magnetic fields has an important effect on the distribution of stellar masses as well as density perturbations
which will be discussed in more detail in section \ref{sect3}.

In general magnetic fields in spiral galaxies have a regular component and a random component. 
Depending on the location within the galaxy the regular or the random component can be dominant.
Typically the field strength of the regular component in spiral galaxies is  of the order 1-5 $\mu$G \cite{beck}.
There are examples of galaxies with much stronger regular magnetic fields of upto 15 $\mu$G such as in the interarm region of NGC 6946 \cite{BeHoe}.
In general, the magnetic field in the spiral arms is dominated by the random component which is due to 
star formation and expansion of supernovae remnants leading to turbulence in the interstellar medium,
thereby entangling the magnetic field lines.
The total magnetic field strengths, including the regular and random components, is on average 
of the order of 9 $\mu$G, however, it in the prominent spiral arms of M51 it is of the order of 30-35 $\mu$G \cite{beck}.

Over recent years the number of Faraday rotation measures $RM$ to estimate the 
magnetic field in our own Galaxy has increased significantly thanks to, e.g.,  the 
National Radio Astronomy Observatory VLA Sky Survey (NVSS) of polarized 
radio sources. In \cite{tss} 37543 $RM$s have been derived. The Galactic magnetic field
has three different components, namely, the central region, the halo and the disk.
The Galactic halo magnetic field in the solar neighbourhood is found to be of  the order of $10^{-1}$ $\mu$G.
The central region is associated with a highly regular magnetic field of strength upto milligauss concentrated 
 in filaments \cite{beck2,han}.
The local regular disk magnetic field is  $4\pm 1$ $\mu$ G and the total magnetic field 
strength $\langle B_t\rangle=6\pm 2$ $\mu$G \cite{beck}.
Multiple field reversals are observed in the Galactic magnetic field.
This is rather unusual and has not been observed in other spiral galaxies. 
However, it might actually be the result of us observing from within the Galaxy, meaning that 
observations trace different volumes, or due to large-scale anisotropic field loops \cite{beck,beck2}.

Observations of clusters of galaxies indicate magnetic fields of the order $\mu$G.
Faraday rotation measurements of the Coma cluster yield a field strength of 2 $\mu$G \cite{kim}.
The structure of the magnetic field depends on the type of cluster. The random component 
is of the order of  $\mu$G for non-cooing flow clusters, such as the Coma cluster with larger correlation length (10-30 kpc) and  
can reach several  $\mu$G for cooling flow clusters such as the Hydra A cluster and shorter correlation lengths \cite{ens}.

There are also observations of magnetic fields associated with high redshift galaxies. In particular in \cite{kron}
Faraday rotation measures of 268 quasars and radio galaxies upto a redshift $z\sim 3.7$ were determined.
It was found that these objects are endowed with $\mu$G-level magnetic fields indicating that they were generated
very quickly at early cosmological epochs.

\section{Generation of primordial magnetic fields}

The wealth of observations of magnetic fields in the universe naturally leads to the question 
of their origin. Clearly the presence of magnetic fields in high redshift objects puts their time of generation
long before the present epoch. Indications of void magnetic fields not associated 
with any gravitationally bound structures add a novel aspect since in this case the generation mechanism cannot rely directly
on, say,  amplification mechanisms during gravitational collapse or dynamo mechanisms.

For galactic magnetic fields, it is generally assumed that a galactic dynamo is operating thereby 
amplifying an initial seed field. Depending on the efficiency of the dynamo the magnetic seed field
can be as small as $B_{seed}=10^{-20}$ G \cite{tw}. Taking into account the contribution of  dark energy observed in our 
universe significantly relaxes the lower bound on the  field strength to explain the present day
galactic magnetic field of $\mu$G-level, lowering it to $B_{seed}\sim 10^{-30}$ G \cite{dlt}.
The standard mechanism is the $\alpha-\Omega$ mean field dynamo theory.
The key equation is \cite{kz}
\begin{eqnarray}
\frac{\partial\vec{B}}{\partial t}=\nabla\times\left(\vec{v}\times\vec{B}\right)+\frac{\eta c^2}{4\pi}\nabla^2\vec{B}
\label{e1}
\end{eqnarray}
where $\eta$ is the resistivity of the plasma.
 $\nabla^2\vec{B}$ can be approximated by $\vec{B}/L^2$ so that the last term can be written in terms of
 a decay time $t_{decay}$ as $\vec{B}/t_{decay}$.
It turns out that in the Galaxy the last term can be neglected since
for a typical gas temperature of $10^4$ K the resistive term $\eta c^2/4\pi\simeq10^7$ cm$^2$s$^{-1}$ \cite{kz}
implies a decay time of $10^{26}(L/1\; {\rm pc})^2$ years where $L$ is the correlation length of the magnetic field.
Neglecting the resistivity term in equation (\ref{e1}) the mean field $\alpha$-$\Omega$ dynamo is determined
by splitting the velocity as well as the magnetic field in a mean part, indicated by an index 0, and a random part, indicated by a $\delta$.
Ensemble averaging over the fluctuations yields to \cite{kz}
\begin{eqnarray}
\frac{\partial\vec{B}_0}{\partial t}=\nabla\times\left(\vec{v}_0\times\vec{B}_0\right)+\nabla\times\left(\langle\delta\vec{v}\times
\delta\vec{B}\rangle\right).
\end{eqnarray}
Solving the corresponding evolution equation for the magnetic field fluctuation $\delta\vec{B}$ in terms of the turbulent
velocity $\delta\vec{v}$ in the quasilinear expansion leads to \cite{kz}
\begin{eqnarray}
\langle\delta\vec{v}\times\delta\vec{B}\rangle=\alpha\vec{B}_0-\beta\nabla\times\vec{B}_0
\end{eqnarray}
where $\alpha=-\frac{1}{3}\tau_c\langle\delta\vec{v}\cdot\left(\nabla\times\delta\vec{v}\right)\rangle$ where $\tau_c$ is the correlation time. Hence
$\alpha$ is determined by the kinetic helicity of the random velocity field.
$\beta=\frac{1}{3}\tau_c\langle\delta v^2\rangle$ which determines the diffusion which smoothes out the turbulent magnetic field component \cite{kulsrud}. Finally the mean field dynamo equation reads \cite{kz}
\begin{eqnarray}
\frac{\partial\vec{B}_0}{\partial t}=\nabla\times\left(\vec{v}_0\times\vec{B}_0\right)+\nabla\times\left(\alpha\vec{B}_0\right)
+\beta\nabla^2\vec{B}_0.
\end{eqnarray}
Applied to the galactic disk it is appropriate to use cylindrical coordinates and the mean velocity is expressed in terms of the galactic rotation.
The key ingredients for the mean field dynamo to work are differential rotation of the galaxy ($\Omega$)  and turbulent motion ($\alpha$).
There are certain problems with the $\alpha$-$\Omega$ mean field dynamo such as achieving enough turbulence and hence amplification, as discussed, e.g., in \cite{BraSub, kz}.

The problem of cluster magnetic fields is still much less understood than that of galactic magnetic fields. As mentioned above clusters of
galaxies are endowed with  magnetic fields at $\mu$G-level and coherence lengths of order 10 kpc. The source of the cluster magnetic fields
could be, for example, outflows from active galaxies thereby transporting bubbles of magnetized plasma into the intergalactic medium.
Though this could only act as seed fields for the cluster fields and would need further amplification. However, in this case the 
$\alpha$-$\Omega$ mean field dynamo does not work since galaxy clusters are found to have only very weak rotation. Small scale turbulent dynamo action is a possibility though the origin of turbulence in  clusters is still not resolved \cite{BraSub}.

All dynamo mechanisms have in common the need for an initial seed magnetic field. Broadly speaking there are two different classes of generation
mechanisms  \cite{kkt, reviews}. One class of models puts the generation of large scale magnetic fields in the very early 
universe during inflation. Similarly as  density perturbations, which 
seed  large scale structure formation, are generated starting with quantum fluctuations of the inflaton field and subsequent amplification 
on super horizon scales, magnetic fields are generated by amplification of perturbations in the electromagnetic field.
First proposed in \cite{tw} it was immediately clear that within standard electrodynamics the field strength of the generated magnetic fields falls short of
the required minimal magnetic field $B_{seed}\sim 10^{-20}$ G, and even the less stringent bound in the presence of 
a cosmological constant, in order to seed the galactic dynamo. It is common to use the ratio of
magnetic field energy density over photon energy density $r\equiv\frac{\rho_B}{\rho_{\gamma}}$ since for a frozen-in magnetic field
$\rho_B\propto a^{-4}$ where $a$ is the scale factor. Since the photon energy density has the same scaling with $a$, $r$ is a constant in this case. 
$B_s\simeq 10^{-20}$G corresponds to $r\simeq 10^{-37}$ and $B_s\simeq 10^{-30}$ G to $r=10^{-57}$.
For a stochastic magnetic field $r$ is calculated using the energy density stored in the mode with comoving wave number $k$, that is $\rho_{\rm B}=k\frac{d\rho_{\rm B}}{dk}$.
Following \cite{tw} we assume that the energy density stored in the mode with comoving wave length $\lambda$ is of the order of the energy density in a thermal bath at the Gibbons-Hawking temperature of de Sitter space.Thus at first horizon crossing the magnetic energy density is given by 
\begin{eqnarray}
\rho_{\rm B}(a_2)\simeq H^4\simeq \Big(\frac{M^4}{M_P^2}\Big)^2
\label{e5.2}
\end{eqnarray}
where the constant energy density during inflation is given by $M^4$ and $a_2$ is the scale factor at the time when the comoving length scale $\lambda$ was crossing the horizon during inflation.
$M_P$ is the Planck mass.
At the end of inflation this implies 
\begin{eqnarray}
r(a_1)\simeq 10^{-104}\Big(\frac{\lambda}{\rm Mpc}\Big)^{-4}\Big(\frac{M}{T_{\rm RH}}\Big)^{10/3}.
\end{eqnarray}
Thus at a galactic scale, $\lambda=1$ Mpc, and typical values for $M$ and the reheat temperature $T_{\rm RH}$, say 
$M=10^{17}$ GeV and $T_{\rm RH}=10^9$ GeV,
$r$ is of the order of $r\simeq 10^{-80}$ which is much below the required minimal  value
even in the presence of a cosmological constant.

Therefore, in the case of a flat background, it is necessary to go beyond the standard model. The key point 
is to effectively change the amplification on superhorizon scales.
There are different possibilities of modifying the standard four dimensional electromagnetic Lagrangian such as coupling to
curvature terms, coupling to a scalar field or extra dimensions. There are models for which 
cosmologically relevant magnetic fields can be generated during inflation, for an extensive review see \cite{kkt}. 
For open universes standard electrodynamics is sufficient \cite{open} (see however \cite{crit}).
In general, the correlation length is not a problem for magnetic fields generated during inflation but rather the field strength.
This is exactly the opposite in the second class of generation mechanisms of large scale magnetic field. For this type of models magnetic fields
are generated after inflation during some phase transition such as the electroweak or QCD phase transition. The underlying 
idea is charge separation as in a battery mechanism. 
Magnetic fields generated in phase transitions have non vanishing magnetic helicity  enabling inverse cascade processes.
The correlation lengths of magnetic fields generated, e.g., during the electroweak phase transitions would have 
correlation lengths of the order of 1 AU. However, this could be increased by inverse cascade  transferring power 
from smaller to larger scales (cf., e.g., \cite{kkt}).

\section{Imprints of primordial magnetic fields on the cosmic microwave background}
\label{sect3}

Magnetic fields are determined by their energy density and pressure as well their anisotropic stress.
These are additional contributions that have to be taken into account when describing the evolution of the 
perturbations in the primordial plasma which later on in the evolution of the universe provide the inhomogeneities
in the gravitational potential necessary for galaxy formation and large scale structure in general.
Moreover, magnetic fields also change the evolution of the baryon velocity due to the contribution of the Lorentz force term. 
Long before recombination electrons and baryons are tightly coupled with the photons due to Thomson scattering of photons
off free electrons. Thus long before decoupling  the Lorentz term also effects the evolution of the photons. All of this leads
to important effects of primordial magnetic fields present before decoupling on the angular power spectrum of the temperature
anisotropies and polarization of the cosmic microwave background as well as the matter power spectrum \cite{cmb}-\cite{kk}. 

Observations exclude the presence of any homogeneous magnetic field on large scales of present day  field strength larger than $10^{-11}$ G \cite{cmf}.
Therefore assuming a gaussian random field it is completely determined by its two point function, in Fourier space, (e.g. \cite{kk})
\begin{eqnarray}
\langle B_i^*(\vec{k})B_j(\vec{k}')\rangle=\delta_{\vec{k}\vec{k}'}P_S(k)\left(\delta_{ij}-\hat{k}_i\hat{k}_j\right)+\delta_{\vec{k}\vec{k}'}P_A(k)i\epsilon_{ijm}\hat{k}_m,
\end{eqnarray}
where $P_S(k)$ is the power spectrum of the symmetric part related to the magnetic energy density, $P_A(k)$ is the power spectrum of the asymmetric part related to the magnetic helicity and a hat indicates a unit vector.
Typically, for these power spectra a power law is assumed. Moreover, since magnetic fields suffer viscous damping before decoupling \cite{damp}
there is a upper cut-off $k_m$. 
In addition to the parity even modes helical magnetic fields induce odd parity modes in the CMB such as cross correlations between the E and B polarization, as well as between temperature and the polarization B mode \cite{helical,kk}.
Using  data from the CMB experiments  WMAP, QUaD and ACBAR \cite{sl} put an upper limit on a non helical magnetic field of 6.4 nG at a scale of 1 Mpc.

There is a characteristic signature of magnetic fields in the linear matter power spectrum at small scales. The origin is in the Lorentz term in the baryon velocity equation which leads to the evolution of the total matter perturbation $\Delta_m$ \cite{sl,kk0}
\begin{eqnarray}
\ddot{\Delta}_m+{\cal H}\dot{\Delta}_m-\frac{3}{2}{\cal H}^2\Delta_m={\cal H}^2\Omega_{\gamma}\Delta_B-\frac{k^2}{3}\Omega_{\gamma}L
\end{eqnarray}
where ${\cal H}=\dot{a}/a$, $\Delta_B$ is the magnetic energy density in terms of the the photon energy density and $L=\Delta_B-\frac{2}{3}\pi_B$ is the Lorentz term
where $\pi_B$ is the magnetic anisotropic stress. During matter domination on small scales the resulting linear matter power spectrum is then 
behaves as ${\cal P}_{\Delta_m}\propto k^4{\cal P}_L$. Therefore the magnetic field adds power on small scales which is absent in the standard $\Lambda$CDM model
without a magnetic field.

\section{Summary}

There is observational evidence for large scale magnetic fields in the universe on a large range of scales. Over decades, there has been a multitude of detections of magnetic fields in galaxies, including our own Milky Way, and cluster of galaxies. These are typically in the range of $\mu$G. In recent years observations of TeV blazars point towards the presence of a truly cosmologically, that is,  void magnetic field  in the femto Gauss range. Observations of the CMB put an  upper limit in the  nG range.

It is generally assumed that a dynamo mechanism amplifies an initial seed magnetic field to the present day $\mu$G level. The origin of the initial seed magnetic field is 
still an open problem. There is a range of proposed mechanisms taking place during inflation in the very early universe. In a flat universe it requires breaking
conformal invariance of standard electrodynamics. The problem here is to achieve strong enough magnetic fields but the correlation length is not a problem.
On the contrary, magnetic fields generated after inflation, during a phase transition can be very strong however their correlation length is very small since it is limited by the horizon size at the time of generation. This problem could be alleviated by using inverse cascade operating due to the helical nature of the fields.
Magnetic fields present before decoupling have an influence on the CMB as well as the matter power spectrum which will offer new possibilities of constraining a truly cosmological magnetic with future experiments.

\section{Acknowledgements}
I would like to thank the organizers of the 40th EPS Conference on Plasma Physics for the invitation and the EPS for financial support.
Spanish Science Ministry grants FPA2009-10612, FIS2012-30926 and CSD2007-00042 are gratefully acknowledged.

\end{document}